# Gate-defined graphene double quantum dot and excited state spectroscopy

*Xing Lan Liu, Dorothee Hug, Lieven M. K. Vandersypen*

Kavli Institute of Nanoscience, Delft University of Technology, P.O. Box 5046, 2600 GA Delft, The Netherlands

xinglan.liu@tudelft.nl

ABSTRACT. A double quantum dot is formed in a graphene nanoribbon device using three top gates. These gates independently change the number of electrons on each dot and tune the inter-dot coupling. Transport through excited states is observed in the weakly coupled double dot regime. We extract from the measurements all relevant capacitances of the double dot system, as well as the quantized level spacing.

Extensive efforts are made in investigating double quantum dots defined by electrostatic gates in various systems such as a GaAs two-dimensional electron gas [1,2], semiconductor nano-wires [3] and carbon nano-tubes [4,5,6,7], with the motivation of realizing quantum computation schemes based on spins in quantum dots [8]. Graphene is a promising candidate for such applications due to the expected long spin coherence time [9,10], and flexibility in device designs offered by its two-dimensional nature. Accidental double dots formed by disorder were found in graphene nanoribbons [11]. More recently, graphene double dot devices have been realized by etching graphene into two small islands separated by a narrow constriction, where the inter-dot coupling was shown to be tunable by a side gate [12, 13]. However, the tunability was limited partially due to the permanent presence of the constriction.

Here we define a double quantum dot device based on a graphene nanoribbon (GNR) using only local top gates. The device contains three top gates. The rightmost and leftmost top gate control the electron number on the right and left dot, respectively. A middle gate is used to tune the inter-dot coupling. The measurements exhibit familiar double dot characteristics [1]. In addition, when the inter-dot coupling is switched off by the middle gate, we observe excited states of the graphene double dot, which has not been reported before. The design principle used here can be applied for defining single and multiple quantum dots along a GNR with independent gate control over barriers and charges.

The device is fabricated on graphene flakes deposited on a substrate by mechanical exfoliation of natural graphite [14]. The substrate consists of highly p–doped Si, covered with 285 nm thermally grown silicon dioxide. From their optical contrast against the substrate, we conclude that the flakes are single-layer [15, 16]. Three electron beam lithography steps are used for fabricating the devices using PMMA as resist. First the source and drain electrodes are fabricated on selected graphene flakes. We use 5/50 nm thick e-beam evaporated Cr/Au as electrodes. In a second step, we cover the region where the GNR will be with 15 nm thick silicon dioxide using e-beam evaporation followed by lift-off. This $SiO_2$ layer not only acts as the etching mask for the GNR, but also forms part of the dielectric for the top gates. The GNR is then patterned by exposing it to an $O_2$/Ar (1:9) plasma [17] for 15 seconds. Without removing the $SiO_2$ etching mask, three local top gates, G1, G2, and G3 are fabricated in the last step. The gates consist of 5/5/20 nm thick e-beam evaporated $SiO_2$/Ti/Au, where an extra layer of $SiO_2$ is evaporated to ensure reliable top gate operations. Here we present measurements from a device where the GNR is 800 nm long and 20 nm wide. For this device, the middle gate G2 is 40 nm in width, separated by 80 nm from gate G1 and G3 which are both 600 nm wide. Fig. 1 shows a scanning electron microscope image of a similar device.

All measurements are performed in a dilution refrigerator at a base temperature of 50 mK. The electron temperature is around 150 mK. We measure the two-terminal resistance through the top gated

GNR devices by applying a DC voltage bias on the source electrode and measuring the current at the drain electrode. The degenerately doped Si substrate is grounded.

The GNR is intrinsically hole-doped when all gates are at zero voltage. Fig. 2a shows the low bias conductance as a function of G1 and G3 while G2 is fixed at zero voltage. When either of the gate voltages is above 0.5 V, current is suppressed by 3 to 4 orders of magnitude, as the Fermi level enters the transport gap locally under the top gates. The conductance increases again when the voltage on either gate is increased further to above 3 V, where the Fermi level is locally in the conduction band and the electrons that are induced in the GNR contribute to transport. The pinch-off voltage for one gate is nearly independent of the other, indicating little cross-coupling in this configuration. Similarly, Fig. 2b shows the low bias conductance as a function of gate G1 and G2 while G3 is fixed at zero voltage. Current is also suppressed by 3 to 4 orders of magnitude when the applied voltage on G2 is above 1.3 V. The pinch-off voltage of G2 is higher than that of G1 and G3, and shows a mild dependence on $V_{G1}$. The voltage on G2 is increased further up to 4 V, but the ribbon below G2 still does not reach heavily n-doping.

A double quantum dot is formed when the voltages on all three gates are increased to close to pinch-off. Fig. 3a plots the low bias conductance as a function of the voltages on G1 and G3, measured at $V_{G2}$ = 1363 mV. It shows a regular honeycomb pattern characteristic of the charge stability diagram of a double quantum dot [1]. The gates G1 and G3 control the number of holes on dot 1 and 2, respectively. Resonant transport occurs at the triple points. Due to co-tunneling we also measure a finite current along all boundaries of the hexagons. From the size of the hexagons, the peak spacing in G1 and G3 is extracted to be $\Delta V_{G1}$ = 6 mV and $\Delta V_{G3}$ = 5 mV, respectively. Thus the capacitance from dot 1 to gate G1 is $C_{G1} \approx e/\Delta V_{G1}$ = 27 aF, and that from dot 2 to gate G3 is $C_{G3} \approx e/\Delta V_{G1}$ = 32 aF, assuming zero level spacing. The large capacitive coupling to these gates indicates that the dot extends far under the gates. Thus the barriers are likely to be induced by the disorder potential instead of being defined by electrostatic potentials induced by the top gates, similar to earlier work [11, 18, 19]. We estimate from the capacitance values that dot 1 (2) extends to roughly 160 nm under gate G1 (G3) [20]. Since the spacing between G1 (G3) and G2 is 80 nm and the ribbon is 20 nm wide, we then assume that the area $A$ of each dot is around 240 nm by 20 nm. The large capacitive coupling allows G1 and G3 to change the number of carriers on dot 1 and dot 2, respectively. Assuming that holes cross over to electrons at around $V_{G1,G3} \approx 1.3$ V, we roughly estimate that in the voltage configuration of Fig. 3 each dot contains around 150 holes, giving a hole density $n = 3 \times 10^{12}$ cm$^{-2}$.

The splitting between each pairs of triple points is indicative of the coupling between the two dots. For the pair of triple points highlighted by the dashed lines in Fig. 3a, the splitting is $\Delta V_{G1}^m = \Delta V_{G3}^m = 0.9$ mV. When the bias voltage is increased, each triple point grows into a triangle due to inelastic transport [1], as shown in Fig. 3b. From the size of the triangles, we extract conversion factors between gate voltages and energy to be $\alpha_1 = eV_{bias}/\delta V_{G1} \approx 0.4e$ and $\alpha_3 = eV_{bias}/\delta V_{G3} \approx 0.4e$. The charging energy of dot 1 and dot 2 is then $E_{c1} = e^2/C_1 = \alpha_1 e/C_{G1} \approx 2.6$ meV, and $E_{c2} = e^2/C_2 = \alpha_3 e/C_{G3} \approx 2.2$ meV, respectively, where $C_1$ and $C_2$ are the total capacitances of dot 1 and dot 2. Applying a model for purely capacitively coupled double dots [1], we extract the inter-dot coupling capacitance $C_m = C_2 C_{G1} \Delta V_{G1}^m / e = C_1 C_{G3} \Delta V_{G3}^m / e \approx 11$ aF, and the coupling energy $E_m = e^2 / C_m (C_1 C_2 / C_m^2 - 1)^{-1} = 0.4$ meV. Table 1 lists also other capacitance values estimated from the hexagons (level spacing is not taken into account), where $C_{G1-2\ (G3-1)}$ is the cross capacitance between G1 (G3) and dot 2 (dot 1).

A further change of the voltage on the gate G2 changes the inter-dot coupling. When $V_{G2} = 1380$ mV, the inter-dot coupling is practically zero, and the charge stability diagram consists of rectangular cells with overlapping triple points (Fig. 3c). Fig. 3d shows high-bias measurements in the same regime, where the pairs of triangles also overlap as a result of the small inter-dot coupling. Resonant transport through excited states is clearly visible in every triple point (the excited states are discussed further below). In this regime, we extract the energy conversion factors to gates G1 and G3 as $\alpha_1 = \alpha_3 \approx 0.6e$. The charging energies of the two dots are $E_{c1} = 3.6$ meV and $E_{c2} = 2.7$ meV, much larger than in the previous regime of strong inter-dot coupling. Other capacitance values estimated from the stability diagram are also listed in Table 1, where a level spacing value of 0.5 meV is now included in the estimate.

We now discuss in detail the excited-state patterns. Fig. 4a and 4b show high resolution measurements of the pair of overlapping triple points enclosed by the dashed line in Fig. 3d at different bias voltages. Along the baseline of the triangle the ground states of the two dots are aligned, and at the center of the baseline (point *d*), they lie exactly in the middle of the bias window, as illustrated by the level scheme in Fig. 4d. At positive bias (Fig. 4a), moving along a line from point *d* to the tip of the triangle (the detuning axis), the energy levels in dot 1 shift upwards while those in dot 2 shift downwards. At point *e*, the ground state of dot 1 aligns exactly with the 1st excited state of dot 2 (Fig. 4e), and resonant transport occurs. The non-resonant background current level is caused by inelastic processes. From these data, the level spacing of dot 2 is extracted to be around 0.6 meV in this charge configuration. At large negative bias, resonant lines parallel to the baseline are also observed (Fig. 4b) due to resonant transport through the ground state of dot 2 and excited states of dot 1. The level spacing of dot 1 is around 0.6 – 0.9 meV. The measured level spacing is comparable to the average level spacing estimated using $\delta E \approx 1/(D(E_F) A) \approx 0.7$ meV, where $D(E_F)$ is the density of states per unit area for 2D graphene at the Fermi energy and is calculated based on the hole density in the dot estimated earlier. For this triple point, the peak current levels of the excited state lines are slightly higher than that of the baseline, likely because the excited states are better coupled to the source / drain contacts [1,7]. In Fig. 4c, we show another pair of overlapping triple points measured at a slightly different charge configuration, where the resonant current through the ground state and excited state are nearly equal.

Transport through the excited states can be analyzed more quantitatively using the result from Stoof and Nazarov for resonant tunneling [1,21]. In the limit of weak inter-dot tunnel coupling $t_m \ll \Gamma_{i,o}$, where $t_m$ is the inter-dot tunnel coupling, and $\Gamma_{i,o}$ are incoming and outgoing tunnel rates, the current *I* follows a Lorentzian line shape as a function of detuning $\varepsilon$, $I(\varepsilon) = (4et_m^2/\Gamma_o) / (1+ (2\varepsilon/h\Gamma_o)^2)$, with *h* the Plank constant. Fig. 4f plots line-cuts along the detuning axis for both positive and negative bias, and Lorentzian line fits to the ground state lines. The fitting is done for the data points outside of the bias triangle in order to minimize the contributions from the inelastic transport [22]. We extract from the fittings a tunnel rate from dot 1 to the drain $h\Gamma_1 \sim 350$ μeV, from dot 2 to the source $h\Gamma_3 \sim 280$ μeV, and an inter-dot tunnel rate $ht_m \sim 10$ μeV. We note that the ground state resonance lines overlap partially with the excited state lines because the tunnel rates to the leads are comparable to the level spacing. This overlap is not taken into account for the fit.

The inter-dot coupling changes non-monotonously as a function of $V_{G2}$, similar to ref. 12. Fig. 5 shows additional data on the evolution of the pair of triple points in Fig. 3a and b, as the voltage on G2 is changed in small steps. Clearly, the splitting between the triple points changes as $V_{G2}$ is varied. This is also shown in Fig. 5h where the inter-dot coupling energy $E_m$ extracted from the data is plotted as a function of $V_{G2}$. The coupling energy can be tuned from around 0.7 meV down to virtually zero. However, the oscillating behavior suggests that most likely, the change of inter-dot coupling is partially due to resonances induced by disorder close to gate G2.

There is likely to be disorder close to gates G1 and G3 as well. This could be the reason for the observation that even when the voltage on G2 is fixed, the vertex splittings and the peak conductance of the vertices vary for different $V_{G1}$ or $V_{G3}$ (Fig. 3 and Fig. 5). It indicates that the gates G1 and G3 also change the dot-to-lead couplings in a similar non-monotonous manner as G2, in addition to controlling the number of carriers on each dot. We also tried to form a single dot through the ribbon by lowering the voltage on the gate G2 to close to zero, but the device could not be tuned to a regime where a well-defined single dot is formed [20], mainly due to strong disorder. At present, disorder thus substantially limits the control over our device. We note however that this is not intrinsic to the device design.

In conclusion, a double quantum dot device is realized in a graphene nanoribbon with multiple top gates. Resonant transport through excited states is observed. The inter-dot coupling strength is tunable over a wide range by the middle gate, although the coupling changes non-monotonously with gate voltage as a result of disorder. Therefore, further progress is needed in order to suppress the influence of disorder on the tunability of the device. We also anticipate that adding two more gates to the left and right dots would allow one to control the number of charges on the two dots and the barriers to the leads separately, which would further improve the controllability of a graphene double dot device. The device demonstrated here represents an important step towards the manipulations of single charges and spins in graphene quantum dots.

ACKNOWLEDGEMENT. We thank Georg Goetz for useful discussions. This work is supported by the Dutch Foundation for Fundamental Research on Matter (FOM).

SUPPORTING INFORMATION AVAILABLE. Estimate of dot area and single dot data where disorder modifies the dot potential.

**Figure 1.** Scanning electron microscope image of a device similar to the one that is measured in this work (scale bar 400 nm). The dashed lines outline the graphene nanoribbon and the dotted lines indicate dot 1 and dot 2.

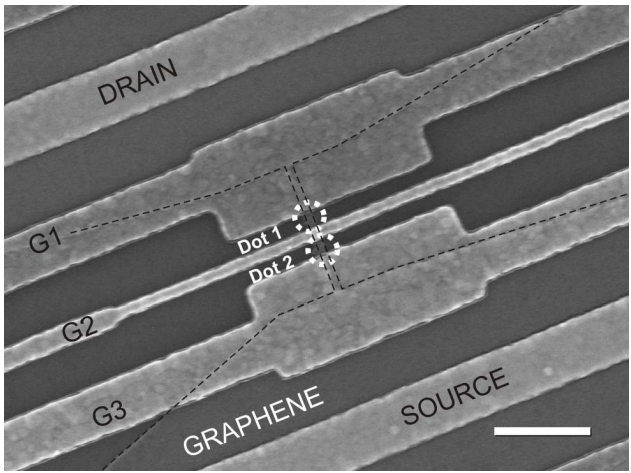

**Figure 2.** Device characterizations (a) Current as a function of top gate voltages $V_{G1}$ and $V_{G3}$ at $V_{G2} = 0$ and $V_{bias} = 100$ μV. (b) Current as a function of top gate voltages $V_{G1}$ and $V_{G2}$ at $V_{G3} = 0$, and $V_{bias} = 300$ μV.

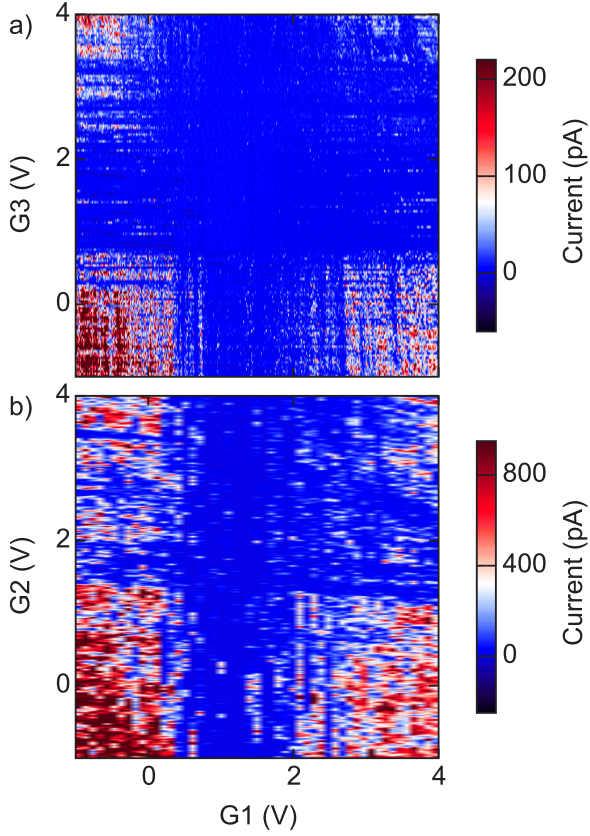

**Figure 3.** Current as a function of top gate voltages $V_{G1}$ and $V_{G3}$ (charge stability diagrams) in the double dot regime at (a) $V_{G2}$ = 1363 mV and $V_{bias}$ = − 15 µV; (b) $V_{G2}$ = 1363 mV and $V_{bias}$ = 0.7 mV; (c) $V_{G2}$ = 1380 mV and $V_{bias}$ = − 20 µV; (d) $V_{G2}$ = 1380 mV and $V_{bias}$ = 1.35 mV. Color scales represent the absolute value of current through the double dot. The white (a) and green (b) dotted lines are guides to the eye showing the honeycomb patterns and the bias triangles. The relevant parameters are also illustrated in (a) and (b). The dashed line in (d) encloses the triple points where the measurements in Fig. 4a, 4b and 5 are taken. The two horizontal shifts at $V_{G1}$ = 644.5 mV and 631.4 mV in (b) are due to charge switching events.

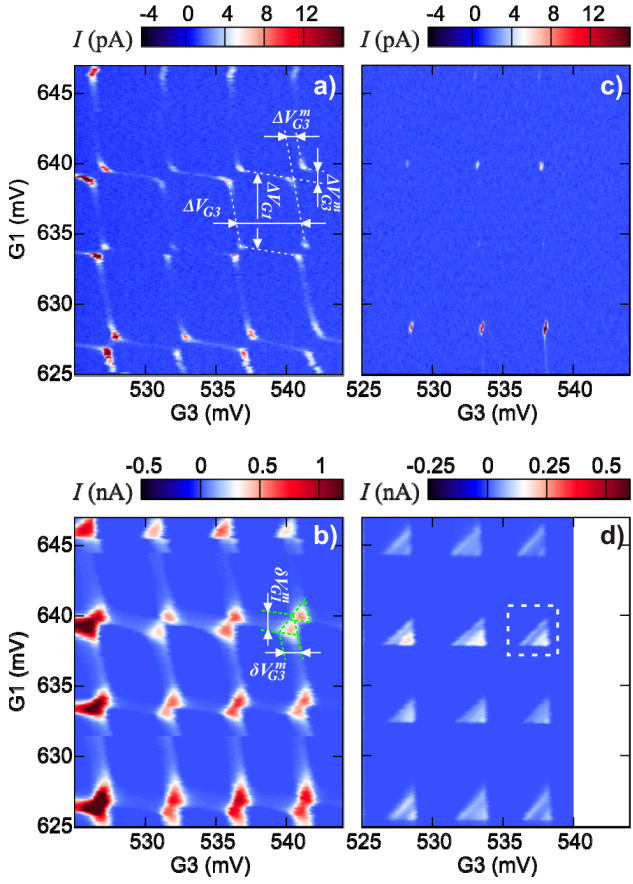

**Figure 4.** Resonant transport through excited states in the double dot. Current as a function of top gate voltages $V_{G1}$ and $V_{G3}$ at (a) $V_{G2}$ = 1380 mV and $V_{bias}$ = 1.9 mV; (b) $V_{G2}$ = 1380 mV and $V_{bias}$ = -1.9 mV; (c) $V_{G2}$ = 1370 mV and $V_{bias}$ = 2.8 mV. The dotted lines indicate the detuning axis. (d, e) Energy level schemes of the double dot corresponding to the points *d* and *e* in (a). Black solid lines represent the ground states and gray lines represent excited states. The chemical potentials of the source and drain contacts are denoted as $\mu_s$ and $\mu_d$, respectively. (f) Line cuts along the detuning axis. The black diamonds, blue circles, and green squares are line-cuts from (a), (b), and (c), respectively. The red solid lines are Lorentzian fits to the right edges of the ground state lines.

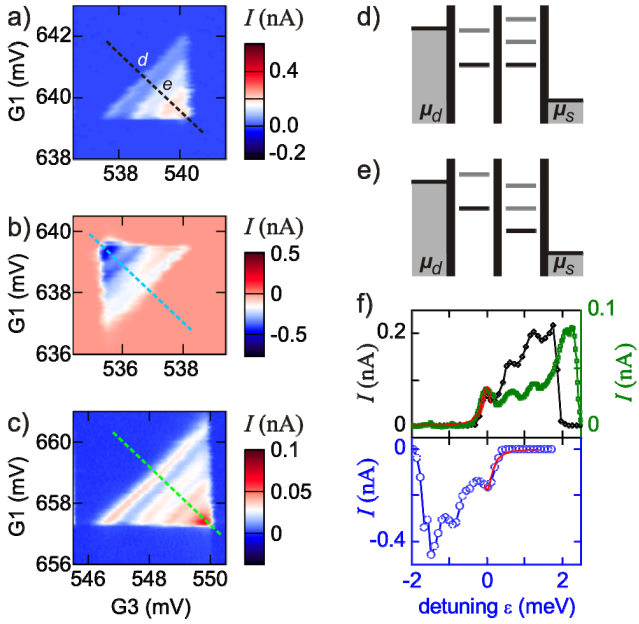

**Figure 5.** Inter-dot coupling vs. middle gate voltage $V_{G2}$. Current as a function of top gate voltages $V_{G1}$ and $V_{G3}$ at $V_{bias} = 15$ μV and (a) $V_{G2} = 1373$ mV, (b) $V_{G2} = 1369$ mV, (c) $V_{G2} = 1367$ mV, (d) $V_{G2} = 1363$ mV, (e) $V_{G2} = 1361$ mV, (f) $V_{G2} = 1359$ mV, (g) $V_{G2} = 1357$ mV, measured at around the same charge configuration as that of Figure 4a. (h) The inter-dot coupling energy $E_m$ as a function of $V_{G2}$ extracted from (a – g) and similar measurements.

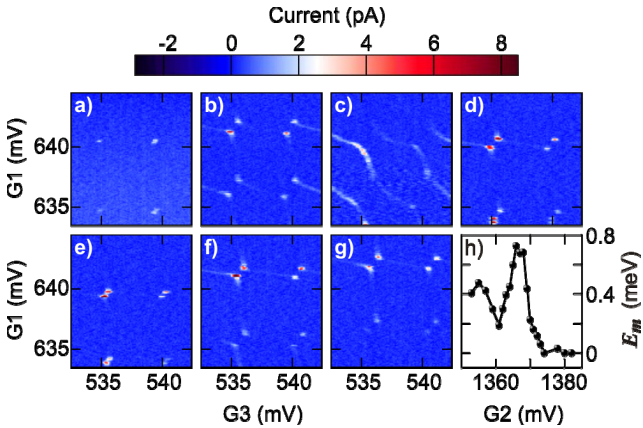

**Table 1.** Capacitance values (in aF) extracted from the honeycomb diagrams shown in Figure 3

| $V_{G2}$ (mV) | $C_{G1}$ | $C_{G3}$ | $C_1$ | $C_2$ | $C_{G1-2}$ | $C_{G3-1}$ |
|---|---|---|---|---|---|---|
| 1363 | 27 | 32 | 59 | 77 | 5 | 6 |
| 1380 | 28 | 38 | 44 | 59 | 0 | 0 |

# Supporting information to "Gate-defined graphene double quantum dot and excited state spectroscopy"

A single quantum dot is expected in the graphene nano-ribbon (GNR) when both G1 and G3 pinch off the GNR while G2 remains hole-doping. Fig. S1c shows the low bias conductance as a function of the voltages on G1 and G3, measured at $V_{G2}$ = 110 mV. We observe resonant lines approximately from top left to bottom right of the plot, indicating the formation of a quantum dot along the GNR. These lines deviate from parallel straight lines expected for a single quantum dot, indicating that the potential landscape in the dot deviates from that of an ideal single dot. We can nevertheless extract the peak spacing in gate G1 and G3 to be $\Delta V_{G1} \approx \Delta V_{G3} \approx 5$ mV. It follows that the capacitances from the dot to the gate G1 and G3 are $C_{G1} \approx C_{G3} \approx 32$ aF, close to the capacitance value in the double dot regime (see main text).

We now estimate roughly how far the dots extend below G1 and G3 in the "single dot" regime from comparison of $C_{G1}$ and $C_{G3}$ to $C_{G2}$. Since the value of $C_{G1}$ and $C_{G3}$ are about the same in the single and double dot regimes, this allows us to also obtain a rough estimate of the dot areas in the double dot regime as well.

Fig. 1a (1b) shows the low bias conductance as a function of the voltages on gates G2 and G1 (G3), measured at $V_{G3}$ = 530 mV ($V_{G1}$ = 640 mV). The peak spacing and the resulting capacitance to gates G1 and G3 remain the same as those from Fig. 1c, as expected. The peak spacing in gate G2 is $\Delta V_{G2} \approx 10$ mV, so the resulting capacitance to gate G2 is $C_{G2} \approx 16$ aF. Given that G2 is 40 nm wide and is separated from the GNR by about 20 nm, and with a GNR width of 20 nm, we estimate that G2 effectively controls the dot potential over an area of 80 nm by 20nm (we hereby make the reasonable assumption that the single dot wave function extends fully across the segment of the GNR below G2). Based on the relative lever arm between the gates, $C_{G1, G3}/C_{G2} \approx 2$, we then estimate the dot area

underneath gates G1 and G3 to be around twice as large, namely, 160 nm by 20 nm. The total area of each of the dots in the double dot regime is then approximately 240 nm by 20 nm.

The dot area can in principle be extracted from the measured $C_{G1,\ G3}$ values and the device geometry based on electrostatic calculations [1]. However, the calculated capacitance between the entire GNR and G2 (for which the geometry is well defined) is $C_{G2}^{cal} = 10$ aF using a dielectric constant of 3.9 for $SiO_2$ and the intended dielectric thickness of 20 nm. This is 1.5 times smaller than the measured capacitance to the dot. It suggests that either the actual gate dielectric is thinner than 20 nm or has a different dielectric constant. This is possible because the thickness of the dielectric from the fabrication is not well calibrated, and the dielectric constant of the evaporated $SiO_2$ is likely to be different from that of the bulk value. An estimate of the dot area based on calculation is thus less applicable.

REFERENCES

1. G.A. Steele, Ph.D. thesis, MIT, **2006**

FIGURE S1. Transport in the single dot regime. (a) Current as a function of top gate voltages $V_{G2}$ and $V_{G1}$ at $V_{G3} = 530$ mV and $V_{bias} = 50$ µV. (b) Current as a function of top gate voltages $V_{G2}$ and $V_{G3}$ at $V_{G1} = 640$ mV and $V_{bias} = 50$ µV. (c) Current as a function of top gate voltages $V_{G1}$ and $V_{G3}$ at $V_{G2} = 110$ mV and $V_{bias} = 50$ µV. (d) Illustration of the dot formation along the GNR in this configuration. The yellow rectangles represent the top gates, and the dashed lines indicate a possible position of the dot.

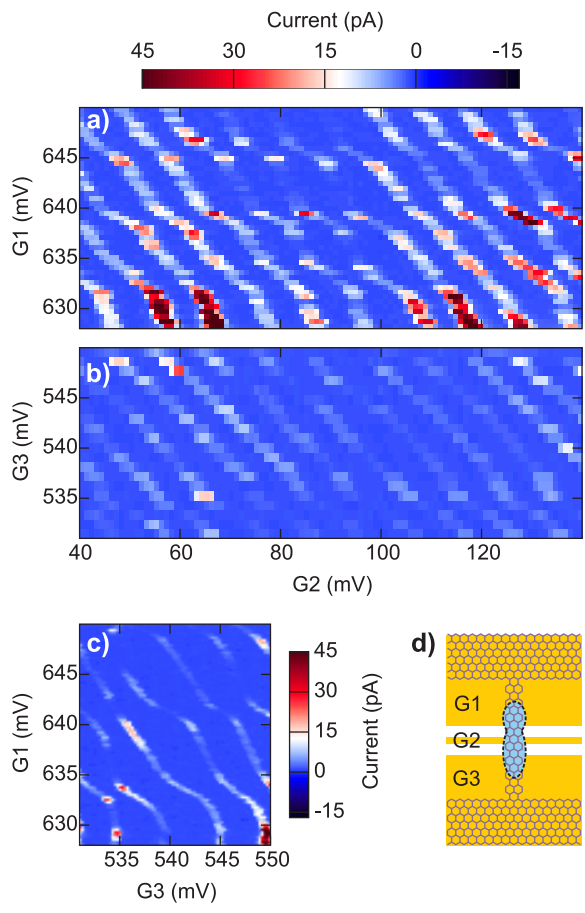